\documentclass{article}
\usepackage[utf8]{inputenc}
\usepackage[british]{babel}
\usepackage{color}
\usepackage{xcolor}
\usepackage{amsmath}
\usepackage{amssymb}
\usepackage{amsthm}
\usepackage{adjustbox}
\usepackage{graphicx}
\usepackage{float}
\usepackage{pdflscape} 

\usepackage{multirow}
\usepackage[mathscr]{eucal}
\usepackage{subcaption}
\usepackage{epstopdf}
\usepackage{caption}
\usepackage{verbatim}
\usepackage{booktabs}
\usepackage{natbib}
\usepackage{colortbl}
 	\definecolor{manatee}{rgb}{0.59, 0.6, 0.67}

\usepackage{rotating}

\usepackage{lineno,hyperref}

\usepackage{ulem}

\begin{document}
\title{Longitudinal Bayesian networks for assessing team performance in the National Basketball Association}

\author{Gabriel Calvo$^{1}$, Francisco Palm\'i-Perales$^{1}$, Carmen Armero$^{1}$,  and \\Virgilio G\'omez-Rubio  $^{2}$}
\maketitle
\noindent $^1$ Department of Statistics and Operations Research,
             Faculty of Mathematics, Universitat de Val\`encia, Spain.
     {\textbf{gabriel.calvo@uv.es}},
             {carmen.armero@uv.es, francisco.palmi@uv.es} \\
$^2$ Department of Mathematics, School of Industrial Engineering-Albacete, Universidad Castilla-La Mancha, Spain.
{virgilio.gomez@uclm.es}

\maketitle

\begin{abstract}
{Assessing the performance of a basketball team requires the consideration of multiple sources of information.} In recent years, the volume and the quality of data generated in sport has increased considerably, particularly in basketball. In this work, we propose a Bayesian graphical modelling framework for the longitudinal analysis of basketball team performance. The framework is based on Bayesian networks that explicitly represent the nodes, that is, the random variables of interest, as longitudinal stochastic processes. We propose three baseline longitudinal models: a static Bayesian network, a dynamic Bayesian network with an autoregressive structure between successive games, and a dynamic Bayesian network based on a hidden Markov structure. We illustrate the proposed framework through a real-world sports analytics case study involving the Philadelphia 76ers of the National Basketball Association (NBA) during the 2005--06 season. {The analysis includes player participation, minutes played, fouls drawn, and one-, two-, and three-point shots attempted and made.}
\end{abstract}

\section{Introduction}

{The analysis of team and player performance in team sports generally requires the joint study of multiple interrelated variables. Bayesian networks (BNs) are particularly well suited to this task because they provide a multivariate modelling framework in which the probabilistic dependencies among the variables can be represented explicitly through a graphical structure. This representation also facilitates the interpretation of the relationships captured by the model.
}

Sports analytics can be defined as the use of statistical, mathematical, and computational methods to analyse sports data and support decision-making concerning player and team performance, strategy, and sports management \citep{alamar2013sports, baumer2023big}. Basketball is one of the most popular team sports in the entire world both at the professional, amateur and collegiate levels. The National Basketball Association (NBA) is an American professional basketball league and the premier professional basketball league in the world.   

  In recent decades, the quality of sport-derived data has increased substantially, particularly in basketball \citep{gudmundsson2017spatio, terner2021modeling}. Such data are well suited to repeated-measures and longitudinal modelling. In repeated-measures designs, multiple observations are obtained from the same experimental units under varying conditions \citep{diggle2002analysis}.

   Bayesian Networks (BNs) are probabilistic models  whose structure can be represented graphically  through   directed acyclic graphs (DAGs). Nodes in BNs are random variables. Edges are probabilistic relationships between nodes that involve real causal relationships.  More specifically, DAGs are directed graphs that contain no directed cycles \citep{Diestel2017}. Within this graphical framework, BNs provide a general approach to assessing model  uncertainty by combining the world of probability and graph theory \citep{Holmes2008}.   As highlighted by \citet{Pearl1995}, the most relevant feature of  BNs  is that they are  ``direct representations of the world, not of reasoning processes''. The history of  BNs is  recent and dates back to the pioneering work of \citet{Pearl1988},   who introduced the probabilistic approach to artificial intelligence and expert systems. Since then, BN have become an indispensable tool in mastering AI in uncertain environments \citep{Wiegerinck2013}.

     In addition, dynamic Bayesian Networks (DBNs) are BNs designed to represent temporal processes \citep{ghahramani1997learning}. The random variables in these type of networks  are indexed by time, and the dependencies are modelled with directed arcs from past to future, usually under a first-order Markov assumption (i.e., each state depends only on the immediately preceding state) or under  autoregressive relationships. This framework generalises classical models such as hidden Markov models \citep{kulkarni2011introduction}, allowing the representation of richer temporal structures with both latent and observable variables.



Although BNs have attracted increasing attention in sports analytics, the applied literature remains sparse. Among the existing contributions, a copula-based Bayesian network jointly models multivariate indicators of players’ performances, capturing non-linear dependencies for interpretable probabilistic comparison of profiles \citep{d2023bayesian}. From a team perspective, structure learning has been used to discriminate `winning' from `losing' team formations and to identify influential variables along the path to victory \citep{zhang2020bayesian}. Regarding temporal scoring patterns, the hot-hand phenomenon has been examined with hidden Markov models, which are particular instances of dynamic BNs, with latent temporal states and observable nodes, thus enabling inference on state-dependent efficiencies and streakiness over time \citep{sun2004detecting,sandri2020markov,otting2022regularized,calvo2025can}. Applications of BNs can also be found in the prediction of the outcomes of football matches \citep{Constantinou_2012, Owramipur_2013, Razali_2017}.

In this context, this article focuses on the analysis of longitudinal basketball data using BNs. We present three foundational models for evaluating basketball player performance over the course of a season through the joint analysis of minutes played, fouls drawn, and shots attempted and made. Utilising longitudinal Bayesian networks, specifically static, dynamic, and dynamic with hidden states, these models are examined through a Bayesian inferential and predictive framework. We illustrate  our methodology using a case study of the Philadelphia 76ers (NBA, 2005–06 season), though the modelling is broadly applicable to any team, league, or season. One of the objectives of the study is to predict the number of points scored by each player in each game. Both model estimation and prediction are carried out within a fully Bayesian framework.

The remainder of the paper is organised as follows. Section 2 outlines the performance data for the Philadelphia 76ers during the 2005–2006 NBA season, which serves as the empirical basis for the analysis conducted in Section 5. Section 3 presents the general Bayesian framework for estimation and prediction in longitudinal Bayesian networks. Section 4 describes the three specific modelling approaches proposed for analysing  basketball data. Section 5 reports  the results obtained from applying the three models to the Philadelphia 76ers 2005-2006 NBA season. Finally, Section 6 discusses the main findings and limitations of the study, presents some conclusions, and outlines directions for future research.

 \section{The Philadelphia 76ers}

The Philadelphia 76ers are a professional basketball franchise based in the Philadelphia metropolitan area, competing in the Atlantic Division of the Eastern Conference within the National Basketball Association (NBA) in the USA. During the 2005–2006 NBA season, this team played a total of 82 games (the complete regular-season schedule), with 38 wins and 44 losses. The team played no additional postseason games, as they finished 9th in the Eastern Conference and failed to qualify for the playoffs.

 The play-by-play data were obtained from NBAstuffer
\citep{nbastuffer_playbyplay} and were accessed on 3 May 2022. The study dataset contains information on the performance of 13 players  of the Philadelphia 76ers team who most frequently participated  during the 2005-2006  regular season, each identified by their playing position (Pos),    guard (PG, SG), forward (SF, PF) and center (C). 

Table \ref{tab:player_stats} presents a summary of individual player statistics for the team during that regular season. It includes the total minutes played (MP), number of personal fouls received, free throw attempts (FT Att), two-point field goal attempts (2PT Att), and three-point field goal attempts (3PT Att), as well as successful free throws (FT Made), two-point field goals (2PT Made), and three-point field goals (3PT Made). These performance metrics provide the foundational data for our Bayesian statistical modelling and analysis of shooting efficiency and player behaviour. Is it worth mentioning that the information in the table is aggregated, with the total for the entire season reported for each player. In this study, however, we analyse the data at the player-game level and adopt a longitudinal modelling structure to account for changes in player performance over the course of the season.

\begin{table}[H]
\centering
\footnotesize
\caption{Player performance summary of the 2005--2006 NBA regular season of the Philadelphia 76ers team: position (Pos), minutes played (MP), fouls received (FR), and free throws, two-pointers and three-pointers  attempted and made shooting statistics.}
\label{tab:player_stats}
\vspace{0.25cm}

\begin{tabular}{lccccccc}
\toprule
Player & Pos & MP & FR & FT Att/Made & 2PT Att/Made & 3PT Att/Made  \\
\midrule
Allen Iverson     & PG & 3103.35 & 672 & 829/675 & 1599/743 & 223/72  \\
Andre Iguodala    & PG & 3085.30 & 258 & 349/263 & 530/ 288 & 158/56    \\
Chris Webber      & PF & 2892.90 & 255 & 348/263 & 1345/596 & 77/21  \\
John Salmons      & SG & 2059.35 & 175 & 200/155 & 434/193  & 87/26  \\
Kevin Ollie       & SG & 1069.02 & 56  & 49/41   & 171/74  & 3/1 \\
Kyle Korver       & SF & 2562.80 & 124 & 119/101 & 323/143  & 438/184  \\
Lee Nailon        & PF & 237.50  & 13  & 15/13 & 80/40   & 0/0\\
Lou Williams      & PG & 144.97  & 9   & 13/8  & 43/21   & 9/2 \\
Matt Barnes       & PF & 538.35  & 33  & 43/29  & 99/57   & 11/2 \\
Michael Bradley   & PF & 368.03  & 9   & 3/2   & 74/31   & 5/1\\
Samuel Dalembert  & C  & 1761.70 & 114 & 132/93 & 368/196  & 1/0  \\
Shavlik Randolph  & PF & 486.38  & 58  & 71/43  & 97/44   & 0/0  \\
Steven Hunter     & C  & 1310.62 & 107 & 138/71 & 291/175  & 0/0   \\
\bottomrule
\end{tabular}
\end{table}

A wide variety of performance profiles can be observed, ranging  from high-
volume shooters such as Allen Iverson to lower-usage players. In particular, Allen Iverson and Andre Iguodala were the most active players in terms of minutes played, contributing significantly to the offensive output of the team, as evidenced by their high number of attempts at shots in all categories. In contrast, players like Lou Williams and Lee Nailon had limited playing time and shot volume, representing more marginal roles. The table also highlights positional diversity, offering insight into how shooting responsibilities and foul patterns vary by position.

\section{Bayesian framework for longitudinal Bayesian networks}


Let $(\boldsymbol{Y}, \boldsymbol{Z})= \{(\boldsymbol{Y}_j, \boldsymbol{Z}_j), \,j=1,\ldots, J\}$ be a longitudinal random vector whose component relating to occasion or time $j$  is in turn   a random vector determined by $V$ observable random variables  and $W$ hidden random variables,  $(\boldsymbol{Y}_j, \boldsymbol{Z}_j)=(Y_j^{(1)}, \ldots, Y_j^{(V)}, Z_j^{(1)}, \ldots, Z_j^{(W)})$. 
We assume that $\boldsymbol \phi$ is a set of random effects and 
$\boldsymbol \theta$ is a set of parameters and hyperparameters that governs the stochastic behaviour of   $(\boldsymbol{Y}, \boldsymbol{Z})$.

A longitudinal Bayesian network (LBN) with nodes $(\boldsymbol{Y}, \boldsymbol{Z}, \boldsymbol \phi, \boldsymbol \theta)$ is a joint probability distribution  for this random vector as follows:  
\begin{align}
  p(\boldsymbol y, \boldsymbol z, \boldsymbol \phi, \boldsymbol \theta)= & \, p(\boldsymbol y, \boldsymbol z \mid \boldsymbol \phi, \boldsymbol \theta)\, p(\boldsymbol \phi \mid \boldsymbol \theta) \,\pi(\boldsymbol \theta),
  \label{eqn:BN_def}
 \end{align}
 \noindent where $ p(\boldsymbol y, \boldsymbol z \mid \boldsymbol \phi, \boldsymbol \theta)$ is the joint conditional probability distribution for $\boldsymbol Y$ and $\boldsymbol Z$ given $\boldsymbol \phi$ and $\boldsymbol \theta$ that represents the LBN sampling model,  $p(\boldsymbol \phi \mid \boldsymbol \theta)$     the conditional distribution of the vector of random effects $\boldsymbol \phi$ given $\boldsymbol \theta$, and 
 $\pi(\boldsymbol \theta)$  a prior distribution for the  vector of parameters and hyperparameters $\boldsymbol \theta$. Note that we are using lowercase letters as variables of a function, in this case, density functions. Later, we will also use lowercase letters to denote the data, without making a distinction, as the context is sufficiently clear to avoid ambiguities.

 LBNs have some important features that should be highlighted. Define the parents $pa(Y_{j}^{(v)})$ ($pa(Z_{j}^{(w)})$) of the random variable $Y_{j}^{(v)}$ ($Z_{j}^{(w)}$) in the network as the set of nodes in $(\boldsymbol Y,\boldsymbol Z)$  whose directed edges point towards 
$Y_{j}^{(v)}$ $(Z_{j}^{(w)})$. If the node $Y_{j}^{(v)}$ ($Z_{j}^{(w)}$) has no parents in $(\boldsymbol Y, \boldsymbol Z)$, its random behaviour is described by the conditional distribution, $p(y_{j}^{(v)} \mid \boldsymbol{\phi}, \boldsymbol{\theta})$ ($p(z_{j}^{(w)} \mid \boldsymbol{\phi}, \boldsymbol{\theta})$). For the sake of simplicity, we assume that the parents $pa(Z_{j}^{(w)})$ of a latent random variable $Z_{j}^{(w)}$ include only hidden nodes from $\boldsymbol Z$. This is not the case for $pa(Y_{j}^{(w)})$ nodes, which can have observable or latent parents. 

Although formally, $\boldsymbol \theta$ and  $\boldsymbol \phi$  are generally parents of the variables in 
$(\boldsymbol Y, \boldsymbol Z)$, we will not include them as members in the subsequent set $pa(\cdot)$  in order to clearly distinguish the nature of the conditional probability distributions with which we will address in each part of the paper.  Note that our approach naturally allows for the inclusion of shared random effects and shared parameters across longitudinal nodes. This induces an additional association structure between longitudinal nodes that is not encoded by the conditional independence relations of the DAG alone.

The global semantics property of the BNs (Pearl and Russel, 2003) allows   the joint distribution   $p(\boldsymbol y, \boldsymbol z \mid \boldsymbol \phi, \boldsymbol \theta)$ to be expressed through the product of conditional distributions on the parents:
\begin{align}
 p(\boldsymbol y, &\boldsymbol z \mid \boldsymbol \phi, \boldsymbol \theta) = \nonumber\\  & \mbox{$\prod$}_{j=1}^J \,    \mbox{$\prod$}_{v=1}^V \, \mbox{$\prod$}_{w=1}^W \,p(y^{(v)}_j \mid pa(y^{(v)}_j), \boldsymbol \phi, \boldsymbol \theta ) \,p(z^{(w)}_j \mid pa(z^{(w)}_j), \boldsymbol \phi, \boldsymbol \theta ).
\label{eqn:chain2}
\end{align}

\noindent In this context, each local conditional distribution,
$p(y^{(v)}_j \mid pa(y^{(v)}_j), \boldsymbol{\phi}, \boldsymbol{\theta})$,
associated with a node and its parent set in the BN can be interpreted as a longitudinal submodel incorporating all relevant components, including covariates.

 {For a node $Y_{j}^{(v)}$ (or $Z_{j}^{(w)}$) in measurement or time $j$, if any of its parent nodes include nodes from previous measurements or times $j-1, j-2, \ldots, 1$, then the model falls within the framework of dynamic BNs, and, in particular, within longitudinal dynamic BNs (LDBNs). In the case of dynamic BNs, the probabilistic relationship between a node and its parents  can be modelled in several ways, for instance, through   {Markovian} or {autoregressives} relationships.}

{Bayesian  learning in LBNs can be performed  following the general Bayesian inferential 
protocol. We always start  by eliciting a prior distribution $\pi(\boldsymbol \theta)$ on $\boldsymbol \theta$. This distribution can be very informative if we have a lot of expert information on $\boldsymbol \theta$ or very vague if we do not have it or have it but do not want to include it in the statistical analysis. A common choice for the distribution of random effects $\boldsymbol \phi $ given the hyperparameters is a multivariate normal distribution centered on zero, with a more or less complex structure in the associated variance-covariance matrix. }

Statistics' favourite food is data. In the case of  LBNs, we considered balanced data  from $N$ individuals from paired  observations of the observable random variables $\boldsymbol Y$ of the network, 
  $\mathcal D=\{y_{ij}^{(v)}, i=1, \ldots, N,\ j=1, \ldots, J,\ v=1,\ldots, V\}$,  where $y_{ij}^{(v)}$ is the observed value of $Y_{j}^{(v)}$ for individual $i$. Prior and experimental information are combined through Bayes' theorem to obtain    the posterior distribution of $(\boldsymbol \phi, \boldsymbol \theta)$, 
\begin{equation}
\pi(\boldsymbol \phi, \boldsymbol \theta \mid \mathcal D) \propto L(\boldsymbol \phi,\boldsymbol \theta)\,  \,p(\boldsymbol \phi \mid \boldsymbol \theta) \, \pi(\boldsymbol \theta),
\end{equation}

\noindent where $L(\boldsymbol \phi, \boldsymbol \theta)$ is the likelihood function of $(\boldsymbol \phi, \boldsymbol \theta)$ for the data $\mathcal D$.

 The posterior distribution $\pi(\boldsymbol \phi,\boldsymbol \theta \mid \mathcal D)$ is usually not analytical, so computational methods must be used to obtain approximate samples from it. Although the field of Bayesian computation has experienced great growth in recent times, MCMC and importance sampling procedures continue to be the ones most popular and widely used \citep{Chib2001}.  The computation of  $\pi(\boldsymbol \phi,\boldsymbol \theta \mid \mathcal D)$  via simulation allows one to approximate the posterior distribution of any relevant   $(\boldsymbol \phi, \boldsymbol \theta)$-dependent output   as well as the joint posterior predictive distribution of the possible outcomes $\boldsymbol Y_{*}$ of the network’s observable variables for a new individual
  \begin{align*}
  p(\boldsymbol y_{*} \mid \mathcal D)=
  &\mbox{$\int$} \, p(\boldsymbol y_{*}, \boldsymbol z, \boldsymbol \phi, \boldsymbol \theta \mid \mathcal D)\, \, \mbox{d}(\boldsymbol z, \boldsymbol \phi, \boldsymbol \theta) \\
  =&\mbox{$\int$} \, p(\boldsymbol y_{*} \mid \boldsymbol z, \boldsymbol \phi, \boldsymbol \theta)\, p(\boldsymbol z \mid \boldsymbol \phi, \boldsymbol \theta) \,\pi(\boldsymbol \phi, \boldsymbol \theta \mid \mathcal D) \, \mbox{d}(\boldsymbol z, \boldsymbol \phi, \boldsymbol \theta) ,\end{align*}
 \noindent from which any posterior predictive marginal distribution or posterior predictive conditional of interest associated with  observable nodes can be approximated.

\section{Bayesian LBNs for assessing   the effectiveness of players on a basketball team}

 We propose three modelling approaches tailored to analyse the evolution of player performance across successive league matches. The first one consists of a static Bayesian network in which each node is treated as a longitudinal  variable. 
In the second approach, we additionally incorporate temporal dependencies between the observable nodes, leading to a dynamic LBN.
Finally, we introduce   latent nodes in the network that follow  a first-order Markov structure over time, thereby defining a hidden dynamic LBN.

\subsection{A static LBN}

 Consider a set of $i=1,\dots, N$  basketball players from a team participating in a $J$-game league  for which we observe the following random variables:  

\begin{itemize} 
\item Number of personal fouls,  ${Y}_{ij}^{(F)}$,  received by player $i$ in game $j$, $j=1, \ldots, J$. 
\item  Number of free throw attempts, ${Y}_{ij}^{(T_1)}$, field goals attempts, ${Y}_{ij}^{(T_2)}$, and three point field goals attempts, ${Y}_{ij}^{(T_3)}$, made by player $i$ in  game $j$.
\item   Number of free throws, ${Y}_{ij}^{(C_1)}$, field goals, ${Y}_{ij}^{(C_2)}$, and three point field goals, ${Y}_{ij}^{(C_3)}$,  scored by player $i$ in game $j$. 
\end{itemize} 
All the random variables defined above depend on the number of minutes played by player $i$ in match $j$, denoted by ${Y}_{ij}^{(M)}$. Since not all players participate in every game, we introduce an indicator random variable to represent whether or not player $i$ takes part in match $j$ during the season. In particular,  we define the random variable ${Y}_{ij}^{(A)}$ as 1 or 0 depending on whether or nor the player $i$ participates in the match $j$. Obviously, if player $i$ does not play in match $j$ we will have  ${Y}_{ij}^{(M)}={Y}_{ij}^{(F)}=0$  and ${Y}_{ij}^{(C_k)}={Y}_{ij}^{(T_k)}=0$, with $k=1,2,3$. In addition, we included a covariate $H_j$ in the modelling  that indicates if the match $j$ is played at home (${I_H}(j) = 1$) or away (${I_H}(j)= 0$), and a set of indicator covariates that inform about the position of each player on the field. In particular, $I_{CE}(i)$, $I_{PF}(i)$, $I_{SF}(i)$, $I_{PG}(i)$, and $I_{SG}(i)$  will be equal to  1 if player $i$ occupies the  center ($CE$), the power forward ($PF$),  the small forward ($SF$), the point guard $(PG$), and the shooting guard ($SG$) position on the field, respectively, and 0 otherwise. 

\subsubsection{Structure}

{We begin by simplifying the notation in order to clarify the structure of the probabilistic models presented in this paper. Let $\boldsymbol Y^{(C)} = (\boldsymbol{Y}^{(C_1)}, 
\boldsymbol{Y}^{(C_2)}, \boldsymbol{Y}^{(C_3)})$
and
$\boldsymbol Y^{(T)} = (\boldsymbol{Y}^{(T_1)}, \boldsymbol{Y}^{(T_2)}, \boldsymbol{Y}^{(T_3)})$  be the random vector with element $\boldsymbol Y^{(C_k)}=\{Y_{ij}^{(C_k)}, i=1, \ldots, N, \,j=1,\ldots, J\}$  and $\boldsymbol Y^{(T_k)}=\{Y_{ij}^{(T_k)}, i=1, \ldots, N, \,j=1,\ldots, J\}$, $k=1,2,3$, respectively. We proposed a Bayesian LBN for the joint distribution of all the above random quantities  as follows, 
\begin{equation}
\begin{split}
     p(\boldsymbol{y}^{(C)},\, &\boldsymbol{y}^{(T)},   \boldsymbol{y}^{(F)},  \boldsymbol{y}^{(M)},  \boldsymbol{y}^{(A)}, \boldsymbol{\theta}, \boldsymbol{\phi})  = \\
      & p(\boldsymbol{y}^{(C)} \mid \boldsymbol{y}^{(T)}, \boldsymbol{\theta}, \boldsymbol{\phi})  \,
      p(\boldsymbol{y}^{(T)} \mid \boldsymbol{y}^{(F)},  \boldsymbol{y}^{(M)},  \boldsymbol{y}^{(A)}, \boldsymbol{\theta}, \boldsymbol{\phi}) \\
       &\,p(\boldsymbol{y}^{(F)}\mid \boldsymbol{y}^{(M)},  \boldsymbol{y}^{(A)}, \boldsymbol{\theta}, \boldsymbol{\phi}) \,
     p(\boldsymbol{y}^{(M)} \mid \boldsymbol{y}^{(A)}, \boldsymbol{\theta}, \boldsymbol{\phi})  \\
     &\,p(\boldsymbol{y}^{(A)} \mid \boldsymbol{\theta}, \boldsymbol{\phi}) \,  p(\boldsymbol{\phi} \mid \boldsymbol{\theta}) \,
      \pi(\boldsymbol{\theta}),
\end{split}
 \label{eqn:joint2}
\end{equation}
\noindent where $\boldsymbol \phi$ will include all   the random effects of the modelling and $\boldsymbol \theta$ will be the set of parameters and hyperparameters of the model. The joint  distribution  in (\ref{eqn:joint2}) factorizes as the product of the conditional joint distribution  $p(\boldsymbol y{^{(C)}}\mid\boldsymbol{y}^{(T)}, \boldsymbol{\theta}, \boldsymbol{\phi})$  of the vector associated with the baskets scored given the shots attempted, $\boldsymbol \theta$ and $\boldsymbol \phi$; the conditional joint distribution  $p(\boldsymbol{y}^{(T)} \mid \boldsymbol{y}^{(F)}, \boldsymbol{y}^{(M)}, \boldsymbol{y}^{(A)}, \boldsymbol{\theta}, \boldsymbol{\phi})$ of   the shots attempted given   the fouls received,  the minutes played by each player in each game, the presence of the player in each game,   $\boldsymbol \theta$ and $\boldsymbol \phi$; the joint conditional distribution of  $p(\boldsymbol{y}^{(F)}\mid \boldsymbol{y}^{(M)}, \boldsymbol{y}^{(A)}, \boldsymbol{\theta}, \boldsymbol{\phi})$ of fouls received in relation to minutes played, the presence of the player in each game, $\boldsymbol \theta$ and $\boldsymbol \phi$; the conditional distribution $p(\boldsymbol{y}^{(M)} \mid \boldsymbol{y}^{(A)}, \boldsymbol{\theta}, \boldsymbol{\phi})$ of the  minutes played with regard to the participation of the player in the match given $\boldsymbol \theta$ and $\boldsymbol \phi$; the
conditional distribution $p(\boldsymbol{y}^{(A)} \mid \boldsymbol{\theta}, \boldsymbol{\phi})$ indicating whether or not each player has participated in each game given $\boldsymbol \theta$ and $\boldsymbol \phi$; and the conditional distribution $p(\boldsymbol{\phi} \mid \boldsymbol{\theta})$ of the random effects given their hyperparameters in $\boldsymbol \theta$ and the prior distribution $\pi(\boldsymbol{\theta})$ for $\boldsymbol \theta$.}  

\begin{figure}[H]
\begin{center}
\includegraphics[width=4cm]{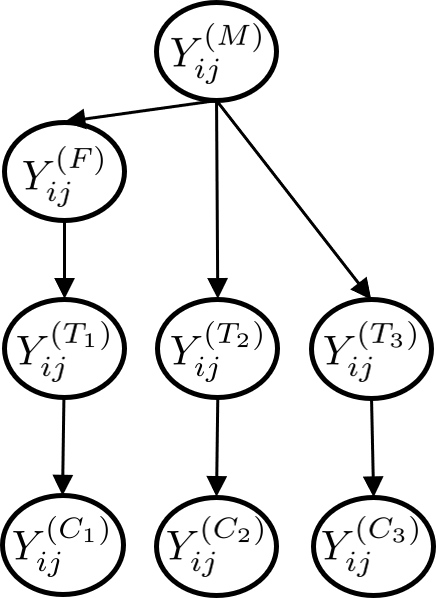} 
\caption{DAG for the scoring LBN of  basketball   player i-th in  the j-th basketball match. For simplicity and better visualisation,  the variable describing each player's possible participation in a match, the vector of parameters, hyperparameters, and random effects have been omitted from the DAG.}
\label{fig:DAG}
\end{center}
\end{figure}

 The DAG for a player $i$ that participates in match $j$ (i.e. $Y_{ij}^{(A)}=1$) associated with this LBN is shown in Figure (\ref{fig:DAG}). The knowledge of $\boldsymbol \theta$ and $\boldsymbol \phi$ assumed by the   conditional distributions of the observables in (\ref{eqn:joint2}) allows the assumption of conditional independence hypotheses  that greatly 
simplify their complete specification. Below we present the proposed modelling for each of these conditional distributions. 
 
\subsubsection{Conditional distributions}

The conditional distributions for the respective network nodes corresponding to games played, minutes played, fouls drawn, shots attempted and shots made are outlined below.

\subsubsection*{Games played}

\noindent {Let $\boldsymbol{Y}^{(A)}$ be the random vector describing the participation of all players in all matches. Its conditional distribution is defined as:
\begin{align}
p(\boldsymbol{y}^{(A)} \mid \boldsymbol{\theta}, \boldsymbol{\phi})= \mbox{$\prod$}_{j=1}^{J}\, \mbox{$\prod$}_{i=1}^{N}\,p(y_{ij}^{(A)} \mid \boldsymbol{\theta}, \boldsymbol{\phi}). 
    \end{align}
    
Recall that $Y_{ij}^{(A)}$ is a binary variable, with probabilities $P(Y_{ij}^{(A)}=1 \mid \boldsymbol{\theta}, \boldsymbol{\phi})$ and $P(Y_{ij}^{(A)}=0 \mid \boldsymbol{\theta}, \boldsymbol{\phi})$,  associated with whether the player $i$ plays or does not play in  match $j$-th of the season. That is, $(Y_{ij}^{(A)} \mid \boldsymbol{\theta}, \boldsymbol{\phi}) \sim \text{Bern}(p^{(A)}_{i})$, where $p^{(A)}_{i}$ denotes the probability that the player $i$ plays in a given match.} Note that, although it could depend on other factors, we are making the simplifying assumption that the probability of playing or not depends only on the player.

\subsubsection*{Minutes played}
The random vector $\boldsymbol{Y}^{(M)}$ representing the minutes played by the basketball players of the team depends directly on $\boldsymbol{Y}^{(A)}$ and ($ \boldsymbol{\theta}, \boldsymbol{\phi}$):
\begin{align}
 p(\boldsymbol{y}^{(M)} \mid \boldsymbol{y}^{(A)}, \boldsymbol{\theta}, \boldsymbol{\phi})= \mbox{$\prod$}_{j=1}^{J}\, \mbox{$\prod$}_{i=1}^{N}\,p(y_{ij}^{(M)} \mid y_{ij}^{(A)}, \boldsymbol{\theta}, \boldsymbol{\phi}).
    \end{align}

We assume a conditional zero-inflated Poisson model \citep{lambert1992zero} (ZIP) for the conditional distribution of $Y_{ij}^{(M)}$ given $Y_{ij}^{(A)}$ and $(\boldsymbol \theta, \boldsymbol \phi)$:
\begin{align}\label{eqn:ZIP}
& (Y_{ij}^{(M)}   \mid Y_{ij}^{(A)}=0,  \boldsymbol \theta, \boldsymbol \phi) \sim 0,  \nonumber \\
& (Y_{ij}^{(M)}   \mid Y_{ij}^{(A)}=1,  \boldsymbol \theta, \boldsymbol \phi) \sim \mbox{Po}(\mu_{i}^{(M)}),  
\end{align} 
\noindent where log($\mu_i^{(M)}) = \mu^{(M)}_0 + b_{i}^{(M)}$, being $\mu^{(M)}_0$    a common mean among all players in all matches, and $(b_{i}^{(M)} \mid \sigma_M) \sim \text{N}(0, \sigma_M^2)$   a random effect associated with  player $i$.

{\subsubsection*{Fouls drawn}
The personal fouls received by each player in a game depend on whether the player participated in the match and, if so, on the number of minutes played.  The conditional distribution of the number $Y^{(F)}$ personal fouls received by all players in all games is as follows:
 \begin{align}
 p(\boldsymbol{y}^{(F)} \mid \boldsymbol{y}^{(M)}, \boldsymbol{y}^{(A)}, \boldsymbol{\theta}, \boldsymbol{\phi})= \mbox{$\prod$}_{j=1}^{J}\, \mbox{$\prod$}_{i=1}^{N}\,p(y_{ij}^{(F)} \mid y_{ij}^{(M)}, y_{ij}^{(A)}, \boldsymbol{\theta}, \boldsymbol{\phi}).
 \label{eqn:cond_yF}
    \end{align}
In this case, we also chose a ZIP distribution to model the conditional distribution in (\ref{eqn:cond_yF}).  When a basketball player does not play a game, the number of fouls drawn is zero. Therefore, 
\begin{align}
& (Y_{ij}^{(F)}   \mid Y_{ij}^{(A)}=0,   \boldsymbol \theta, \boldsymbol \phi) \sim 0, \nonumber \\
& (Y_{ij}^{(F)}   \mid  y_{ij}^{(M)}, Y_{ij}^{(A)}=1, \boldsymbol \theta, \boldsymbol \phi) \sim \mbox{Po}(\mu_{ij}^{(F)}), \nonumber 
\end{align}
\noindent being log($\mu_{ij}^{(F)}) = \mu_0^{(F)} + \beta_{M}^{(F)} \,y_{ij}^{(M)} + b_{i}^{(F)}$, where $\mu_0^{(F)}$ is a common parameter associated with all basket players in all matches, $(b_{i}^{(F)} \mid \sigma_F) \sim \text{N}(0, \sigma_F^2)$ is a random effect associated with the player $i$, and $\beta_M^{(F)}$ is the regression coefficient associated with the minutes played by each player in each match.}
 
\subsubsection*{Shots attempted}
 
{The shots taken by each basketball player depend on their presence in the match and, if positive, on the number of minutes played.  The exception is   for free throws, which depend on the fouls received or assigned to each player. In this way, the conditional distribution for the number of shots thrown for al players in all games is
\begin{align}
  p(& \boldsymbol{y}^{(T)}\mid  \boldsymbol{y}^{(F)}, \boldsymbol{y}^{(M)},\boldsymbol{y}^{(A)},\boldsymbol{\theta}, \boldsymbol{\phi})  = p((\boldsymbol{y}^{(T_1)}, \boldsymbol{y}^{(T_2)}, \boldsymbol{y}^{(T_3)}) \mid  \boldsymbol{y}^{(F)}, \boldsymbol{y}^{(M)},\boldsymbol{y}^{(A)},\boldsymbol{\theta}, \boldsymbol{\phi}) \nonumber\\
  & = \mbox{$\prod$}_{j=1}^{J}\, \mbox{$\prod$}_{i=1}^{N}\, p(y_{ij}^{(T_1)} \mid  y_{ij}^{(F)},  y_{ij}^{(A)}, \boldsymbol{\theta}, \boldsymbol{\phi}) \, \mbox{$\prod$}_{k=2}^{3}\,p(y_{ij}^{(T_k)} \mid y_{ij}^{(M)}, y_{ij}^{(A)}, \boldsymbol{\theta}, \boldsymbol{\phi}).
\end{align} 
We assume a ZIP distribution for each player's throws in each match. In the case of free throw attempts,
\begin{align}
& (Y_{ij}^{(T_1)}   \mid Y_{ij}^{(A)}=0,   \boldsymbol \theta, \boldsymbol \phi) \sim 0, \nonumber \\
& (Y_{ij}^{(T_1)}   \mid  y_{ij}^{(F)}, Y_{ij}^{(A)}=1, \boldsymbol \theta, \boldsymbol \phi) \sim \mbox{Po}(\mu_{ij}^{(T_1)}). \nonumber 
\end{align}

Assuming the center ($C$)  position of a player  as the reference category, the mean of the free throws for each payer and match
is expressed as
\begin{align}
    \mbox{log}(\mu_{ij}^{(T_1)})  = & \,  \beta_0^{(T_1)}+ \beta_F^{(T_1)} y_{ij}^{(F)} + \beta_{PF}^{(T_1)} \,I_{PF}(i) + \beta_{SF}^{(T_1)} \,I_{SF}(i)  \nonumber\\
    &+ \beta_{PG}^{(T_1)} \,I_{PG}(i)+ \beta_{SG}^{(T_1)} \,I_{SG}(i)+ b_{i}^{(T_1)}, 
    \label{eqn:throws}
\end{align}
 \noindent where $\beta_0^{(T_1)}$ is the regression coefficient associated with the center position of the player,   $\beta_F^{(T_1)}$ a slope associated with the faults received by each player in each game, $I_{Q}(i)$ is an indicator function that takes the value 1 if player $i$ has characteristic $Q$ and zero otherwise, and $\beta_{PF}^{(T_1)}$, $\beta_{SF}^{(T_1)}$,
    $\beta_{PG}^{(T_1)}$, $\beta_{SG}^{(T_1)}  $ are the regression coefficients  associated with each of  the player's position on the field with regard to the center position: point guard  ($PG$), the shooting guard ($SG$), the small forward ($SF$), and the power forward ($PF$).   The random effect associated with the  player $i$ is $(b_i^{(T_1)} \mid \sigma_{T_1}) \sim \mbox{N}(0, \sigma_{T_1}^2)$.}

{Similarly, the numbers of two-point and three-point baskets attempted depend on the minutes played and the unknown parameter vectors $(\boldsymbol \theta, \boldsymbol \phi)$ according to the ZIP distribution 
\begin{align}
& (Y_{ij}^{(T_k)}   \mid Y_{ij}^{(A)}=0,   \boldsymbol \theta, \boldsymbol \phi) \sim 0, \nonumber \\
& (Y_{ij}^{(T_k)}   \mid  y_{ij}^{(M)}, Y_{ij}^{(A)}=1,  \boldsymbol \theta, \boldsymbol \phi) \sim \mbox{Po}(\mu_{ij}^{(T_k)}), \,\,k=2,3, \nonumber 
\end{align}
with 
\begin{align*}
    \mbox{log}(\mu_{ij}^{(T_k)}) = & \,\beta_0^{(T_k)} +  \beta_M^{(T_k)} y_{ij}^{(M)}  + \beta_{PF}^{(T_k)} \,I_{PF}(i) + \beta_{SF}^{(T_k)} \,I_{SF}(i)\\
    &+ \beta_{PG}^{(T_k)} \,I_{PG}(i) + \alpha_{SG}^{(T_k)} \,I_{SG}(i)+b_{i}^{(T_k)},
\end{align*}

\noindent where the $\beta$'s parameters have the same meaning as those defined in (\ref{eqn:throws}) except for $\beta_M^{(T_k)}$,   which is now the coefficient associated with the minutes played by each player and  match, and $(b_{i}^{(T_k)} \mid \sigma_{T_k}) \sim \mbox{N}(0,  \sigma_{T_k}^2)$ is a normally distributed random effect associated with  player $i$. }

\subsubsection*{Shots scored}
Finally, the conditional joint distribution of the random vector that represents the basket scores is given by the attempted shots, $\boldsymbol \theta$ and $\boldsymbol \phi$ as follows:
\begin{align}
    p(\boldsymbol y^{(C)} \mid  \boldsymbol y^{(T)}, \boldsymbol{\theta}, \boldsymbol{\phi}) = \mbox{$\prod$}_{j=1}^{J}\, \mbox{$\prod$}_{i=1}^{N}\, \mbox{$\prod$}_{k=1}^{3}\,p(y_{ij}^{(C_k)} \mid y_{ij}^{(T_k)}, \boldsymbol{\theta}, \boldsymbol{\phi}). 
\end{align}
Each shot that a player throws in a game can generate a basket or not, so the number of baskets scored will be a binomial-type variable with a probability that depends on the characteristics of the player, the match and the position the player occupies on the court. Consequently,
\[
   (Y_{ij}^{(C_k)} \mid y_{ij}^{(T_k)}, \boldsymbol{\theta}, \boldsymbol{\phi}) \sim \text{Bi}(y_{ij}^{(T_k)}, \,p_{ij}^{(C_k)}), \,\, k=1,2,3,
    \]
    with 
    \begin{align} \label{eqn:succ_prob}
    \text{logit}(p_{ij}^{(C_k)}) =  & \,\beta_0^{(C_k)}  + \beta_H^{(C_k)} I_{H}(j) +   \beta_{PF}^{(C_k)}  I_{PF}(i)  + \beta_{SF}^{(C_k)}  I_{SF}(i) \nonumber \\ &  + \beta_{PG}^{(C_k)}  I_{PG}(i)  + \beta_{SG}^{(C_k)}  I_{SG}(i) + b_{i}^{(C_k)},
    \end{align}
    where $\beta^{(C_k)}_0$ is a common baseline parameter associated with each player in the center position and match  played away from home, $b_{i}^{(C_k)}$ is a random effect associated with player $i$, where
    $(b_{i}^{(C_k)} \mid {\sigma_{C_{k}}}) \sim \mbox{N}(0, \sigma_{C_{k}}^2)$. Furthermore, $\beta_H^{(C_k)}$ is the coefficient associated with the indicator $I_H(j)$, which takes the value 1 when the game $j$ is played at home and 0 when it is played away. Finally, $\beta_{PF}^{(C_k)}$, $\beta_{SF}^{(C_k)}$, $\beta_{PG}^{(C_k)}$, and $\beta_{SG}^{(C_k)}$ are the coefficients associated with the indicator variables for the player's court position. The center position is used as the reference category.

\subsubsection{Prior distribution}

We assume prior independence among the parameters and hyperparameters of the model and consider a scenario of poor prior information. We elicit  minimally informative normal distributions with mean \( 0 \) and standard deviation \( 10  \) for all regression coefficients of all conditional distributions  of the DAG and uniform distribution on the interval \( (0, 100) \) for all standard deviation hyperparameters. Additionally, the prior distribution for binomial and  zero-inflation probabilities  are considered as  uniformly distributed between \( 0 \) and \( 1 \).

\subsection{A dynamic LBN} 

{In the previous network, we did not consider any temporal probabilistic relationship between a node and its parents. In the event that this relationship is to be included, one way to construct a dynamic LBN is, as mentioned above, to assume an autoregressive relationship between nodes so that the parents of a node at time (match) $j$ can include nodes from the previous time (match) $j-1$. }

\begin{figure}[H]
\begin{center}
\includegraphics[width=8.5cm]{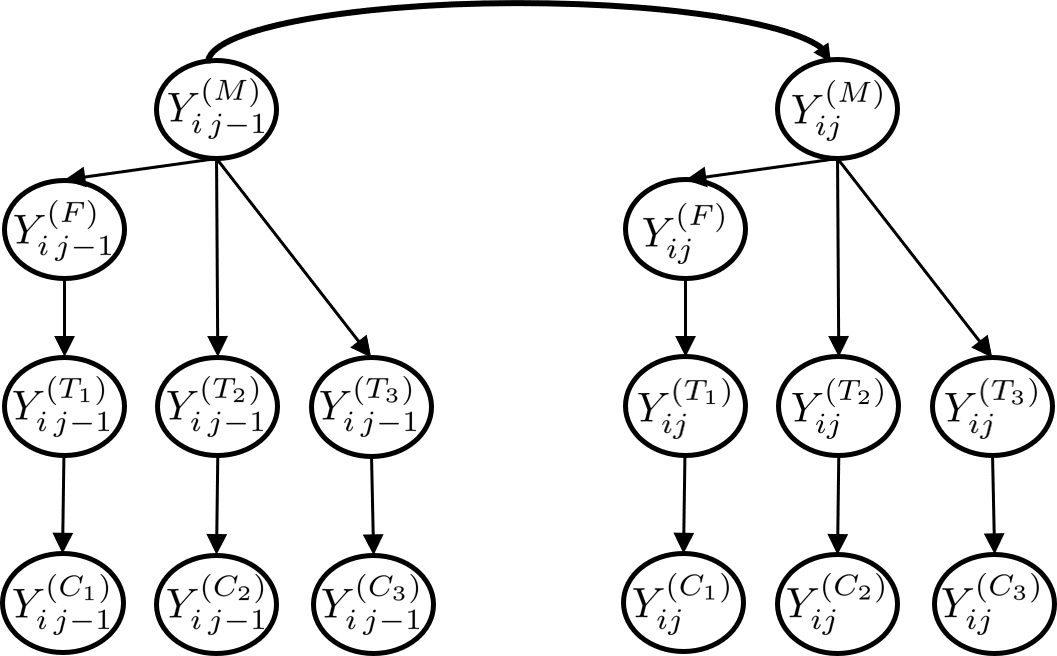} 
\caption{{DAG for the random variables of the Dynamic LBN of a basketball player $i$   who has participated in match $j$. }}
\label{fig:DAG_DBN}
\end{center}
\end{figure}

{In our basketball study, we have represented each match's play through the static network in Figure \ref{fig:DAG}, so a first-order auto-regressive structure can be added by connecting the networks corresponding to two consecutive matches. Figure \ref{fig:DAG_DBN} shows a dynamic network that connects the networks of two consecutive matches through the minutes played by each player. Also, in this DAG, we have omitted the node describing the possible participation of player $i$ in match $j$, and all the parameters, hyperparameters, and random effects of the model. }

{All probabilistic relationships between nodes in this network  remain the same as those presented for the static network in the previous section, except for the distribution of the node describing the minutes played by each player in each match. In this case, the conditional distribution of $\boldsymbol{Y}^{(M)}$ is defined as
\begin{align}
 p(\boldsymbol{y}^{(M)} &  \mid \boldsymbol{y}^{(A)},  \boldsymbol{\theta}, \boldsymbol{\phi})= \nonumber\\   & \, \mbox{$\prod$}_{i=1}^{N}\,p(y_{i1}^{(M)} \mid y_{i1}^{(A)}, \boldsymbol{\theta}, \boldsymbol{\phi})  \mbox{$\prod$}_{j=2}^{J}\, p(y_{ij}^{(M)} \mid y_{i\,j-1}^{(M)}, y_{ij}^{(A)}, \boldsymbol{\theta}, \boldsymbol{\phi}).
    \end{align}}

{For the case $j = 1$, this conditional model is the same as in the static network, that is, 
\begin{align}
& (Y_{i1}^{(M)} \mid Y_{i1}^{(A)} = 0, \boldsymbol{\theta}, \boldsymbol{\phi}) \sim 0, \nonumber \\
& (Y_{i1}^{(M)} \mid Y_{i1}^{(A)} = 1, \boldsymbol{\theta}, \boldsymbol{\phi}) \sim \text{Po}(\mu_{i}^{(M)}), \nonumber
\end{align}}

\noindent with $\log(\mu_{i}^{(M)}) = \mu_0^{(M)} + b_i^{(M)}$. 
The dynamic structure for $j>1$ is introduced following   the approach by \citet{fokianos2011log} 
for log-linear autoregressive time series models:
\begin{align}
& (Y_{ij}^{(M)} \mid Y_{ij}^{(A)} = 0, \boldsymbol{\theta}, \boldsymbol{\phi}) \sim 0, \nonumber \\
& (Y_{ij}^{(M)} \mid y_{i\, j-1}^{(M)}, Y_{ij}^{(A)} = 1, \boldsymbol{\theta}, \boldsymbol{\phi}) \sim \text{Po}(\mu_{ij}^{(M)}), \nonumber
\end{align}
\noindent where
\begin{align}
    \log(\mu_{ij}^{(M)}) = \mu_0^{(M)} +\beta_M^{(+)} \log(1 + y_{i\, j-1}^{(M)})+ b_i^{(M)}.  
\end{align}
Now $\mu_0^{(M)}$ also represents an overall mean, $b_i^{(M)}$ a random effect associated with player $i$, $(b_i \mid \sigma_M) \sim \mbox{N}(0, \sigma_M^2)$, and $\beta^{(+)}_M$ is the coefficient associated with the autoregressive information from $y_{i\, j-1}^{(M)}$ with prior distribution $\pi(\beta^{(+)}_M)=\text{N}(0,10^2)$, following the same elicitation scheme as the previous model.

\subsection{A hidden Markov LBN} 

{We   propose an alternative dynamic model, this time including certain latent nodes. Specifically, we will construct a dynamic Bayesian network with a hidden Markov model  based on the ‘hot hand’ effect  \citep{calvo2025can} which will affect the success of baskets scored. 

We first define a stochastic process that will invisibly govern the behaviour of the network. Let $\{Z_{j}, j = 1,\dots,J\}$ be a hidden Markov chain (HMC) representing the latent state of the team in game $j$, where $Z_j \in \{C,H\}$ indicates whether the team is in a cold ($C$) or a hot ($H$) state. The latent process is assumed to satisfy the Markov property,
\[
P(Z_j \mid Z_1,\dots,Z_{j-1},\boldsymbol{\theta},\boldsymbol{\phi})
=
P(Z_j \mid Z_{j-1},\boldsymbol{\theta},\boldsymbol{\phi}).
\]

The transition probability matrix of the chain is defined as
\begin{equation}
\Omega =
\begin{pmatrix}
q^{CC} & q^{CH} \\
q^{HC} & q^{HH}
\end{pmatrix},
\label{transition1}
\end{equation}

\noindent where $q^{CH}=P(Z_{j}=H \mid Z_{j-1}=C, \boldsymbol{\theta}, \boldsymbol{\phi})$ denotes the transition probability from the cold state $C$ to the hot state $H$ in game $j$, conditional on the model parameters $\boldsymbol{\theta}$ and $\boldsymbol{\phi}$. Finally, consider that $\Omega$ is a stochastic matrix, consequently  each row of the transition matrix sums to one, that is,
$q^{CC}+q^{CH}=1$ and $q^{HC}+q^{HH}=1$.

{The structure of the hidden Markov LBN is shown in  Figure \ref{fig:DAG_HMMDBN}. As we can observe, the probabilistic relationships between player participation and minutes played in successive matches, fouls drawn and shots taken will be the same as those presented for the static network and will not be affected by the hidden process. The baskets scored during a match will depend not only on the shots attempted but also on the latent state of the team in the match as follows:   
\[
p(\boldsymbol y^{(C)} \mid  \boldsymbol y^{(T)}, \boldsymbol z, \boldsymbol{\theta}, \boldsymbol{\phi}) = \prod_{j=1}^{J} \prod_{i=1}^{N} \prod_{k=1}^{3} p(y_{ij}^{(C_k)} \mid y_{ij}^{(T_k)},  z_j, \boldsymbol{\theta}, \boldsymbol{\phi}).
\]
 
 The number of baskets scored $Y_{ij}^{(C_k)}$, $k=1, 2, 3$, will follow  a binomial distribution depending on the baskets thrown  and the current state of the team. In particular, a higher success probability when $Z_j = H$, or a lower one when $Z_j = C$:

\[
\begin{aligned}
&(Y_{ij}^{(C_k)} \mid Z_j=C, y_{ij}^{(T_k)}, \boldsymbol{\theta}, \boldsymbol{\phi}) \sim \text{Bi}(y_{ij}^{(T_k)}, \,p_{C_{ij}}^{(C_k)}), \\
&(Y_{ij}^{(C_k)} \mid Z_j=H, y_{ij}^{(T_k)}, \boldsymbol{\theta}, \boldsymbol{\phi}) \sim \text{Bi}(y_{ij}^{(T_k)}, \,p_{H_{ij}}^{(C_k)}),
\end{aligned}
\]
\noindent where
\[
\begin{aligned}
\text{logit}(p_{C_{ij}}^{(C_k)}) = &\, \beta_{C0}^{(C_k)} +\beta_H^{(C_k)} I_{H}(j) + \beta_{PF}^{(C_k)} I_{PF}(i) + \beta_{SF}^{(C_k)} I_{SF}(i) \\
& + \beta_{PG}^{(C_k)} I_{PG}(i) + \beta_{SG}^{(C_k)} I_{SG}(i) +  b_{i}^{(C_k)} , \\
\text{logit}(p_{H_{ij}}^{(C_k)}) = &\, \beta_{H0}^{(C_k)} + \beta_H^{(C_k)} I_{H}(j) + \beta_{PF}^{(C_k)} I_{PF}(i) + \beta_{SF}^{(C_k)} I_{SF}(i) \\
& + \beta_{PG}^{(C_k)} I_{PG}(i) + \beta_{SG}^{(C_k)} I_{SG}(i)+ b_{i}^{(C_k)}.
\end{aligned}
\] 

\begin{figure}[H]
\begin{center}
\includegraphics[width=11cm]{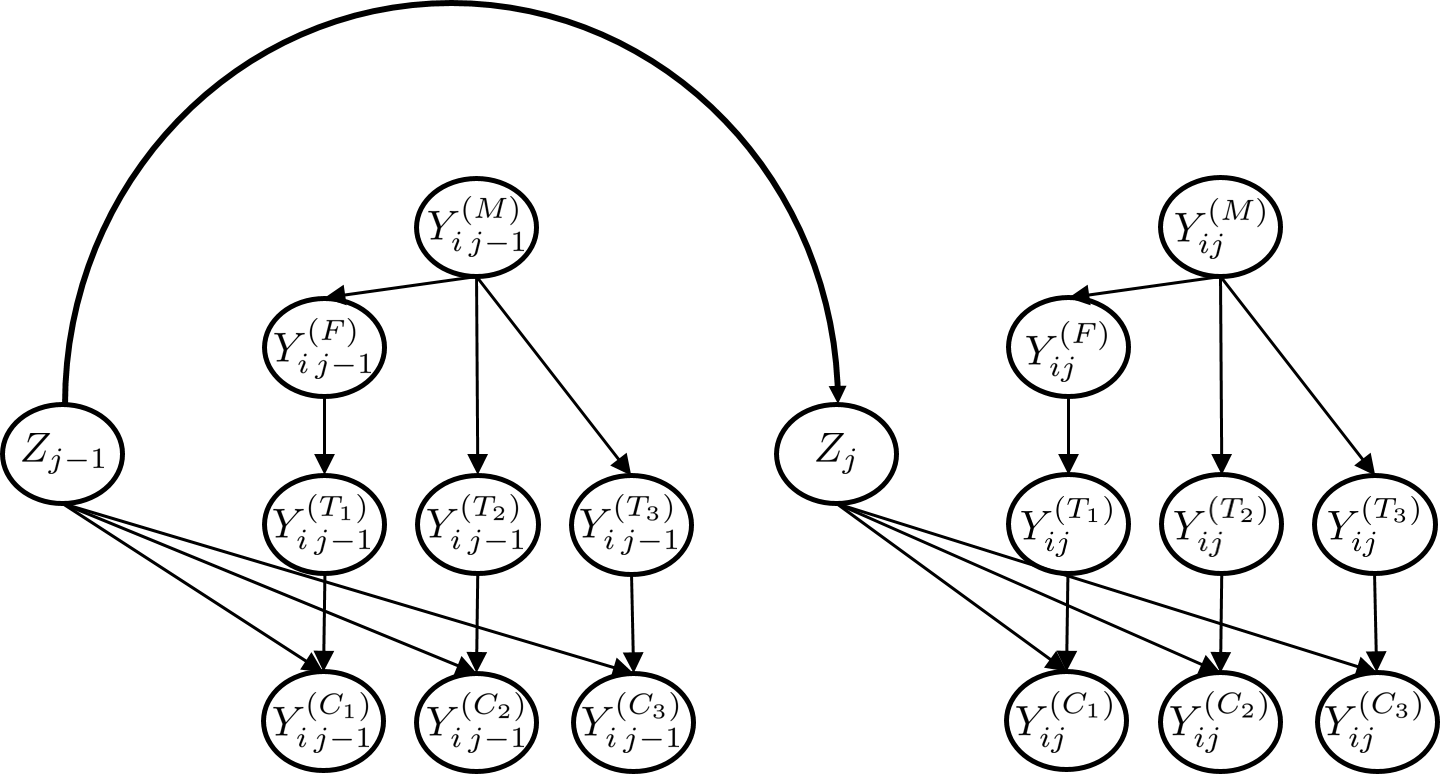} 
\caption{DAG for the hidden Markov dynamic LBN of a basketball player $i$ in  a team basketball match team $j$, when  $Y_{ij}^{(A)}=1$. Nodes for parameters, hyperparameters, and random effects have  omitted in the DAG.  }
\label{fig:DAG_HMMDBN}
\end{center}
\end{figure}

Prior independence,   wide normal distribution centered at 0,  N$(0, 10^2)$ are elicited for the regression coefficients and  beta prior distribution  for the two transition probabilities $q^{CH}$ and $q^{HC}$, $p(q^{CH})=p(q^{HC})=\text{Beta}(1,1)$. It is worth noting that   the restriction $\beta_{C0}^{(C_k)} < \beta_{H0}^{(C_k)}$ is imposed, ensuring that the team’s overall success probability is always higher in the hot state. Moreover, this constraint ensures that we avoid the identifiability problem caused by label switching, which is inherent in models with latent variables. }

\section{Philadelphia 76ers Player Performance}

Inference across all three models for the Philadelphia 76ers performance dataset is based on the joint posterior distribution of the parameters, hyperparameters, and random effects. To obtain an approximation of the corresponding posterior distribution we used the version 1.3.0 of the \texttt{NIMBLE} R package \citep{de2017programming}, which relies on algorithms based on MCMC methods \citep{tanner1993tools}. It is important to highlight the flexibility of \texttt{NIMBLE}, which does not rely on a single MCMC algorithm. Instead, it implements a modular system that combines several samplers, which are automatically selected according to the structure of the model, while always allowing the user to modify them if required.

For each model, we ran three parallel chains with a total of 1,000,000 iterations per chain, including a burn-in period of 500,000 iterations for each chain. To reduce autocorrelation in the samples, the chains were thinned by retaining every 500th iteration. All computations were performed on a computer equipped with an Intel i9 processor, 32 GB of RAM, and running Windows 10 Enterprise LTSC. All code required to reproduce the results in this paper is available at 
\url{https://github.com/gcalvobayarri/Longitudinal_BNs}.

\subsection{Model selection}
We compare the three modelling approaches using the default implementation of the Watanabe Akaike Information Criterion (WAIC) in the \texttt{NIMBLE} environment. The WAIC, introduced by \citet{watanabe2010asymptotic}, is a metric employed for Bayesian model comparison {that makes use of the information of the full posterior distribution via the MCMC sample from it}. It provides an estimate of the out-of-sample predictive error. The underlying idea, shared with other information criteria, is that the model with the smallest value is preferred, as it indicates better expected predictive performance.

Table \ref{tab:waic_comparison} reports the WAIC values computed for the three LBNs. Based on this criterion, the dynamic LBN is the preferred model, as it attains a lower WAIC value than the remaining alternatives. Consequently, from this point forward, we present the results for the dynamic network with an autoregressive structure in the minutes played variable, namely the dynamic LBN.

\begin{table}[H]
    \centering
    \caption{WAIC values for the three modelling approaches.}
\label{tab:waic_comparison}
    \begin{tabular}{lc}
        \toprule
    \textbf{Model} & \textbf{WAIC} \\
        \midrule
        Static LBN  & 21274.69 \\
        Dynamic LBN  & 21194.13 \\
        Hidden Markov LBN & 21274.60 \\
        \bottomrule
    \end{tabular}
 \end{table}

\subsection{Estimation}

{Due to the large number of parameters and hyperparameters involved in the selected dynamic LBN, the posterior distribution of all of them is presented following the ordering of the joint distribution in expression (\ref{eqn:joint2}). We begin with the conditional model for the   successful shots  and conclude with the indicator random variable that assesses the participation of a player in a game.}

{Table \ref{tab:shots_made_new} shows a summary of the approximate  posterior distribution for the parameters and hyperparameters  of the logistic regression models that assess the success of  one-, two- and three-point shots by players in the same basketball match.   Although in many cases we observe rather wide credibility intervals, particularly in the approximate posterior distribution of the parameters associated with the three-point shooting variable, relevant information can still be extracted from the data. For instance, when comparing the standard deviation parameter associated with the players' random effects, we observe that {the standard deviation associated with the probability of success on a 2-point shot for each player,} $\sigma_{C_2}$, exhibits smaller values {whilst the variability in free throws associated with the different players is much greater,  slightly more than double on average.} This could be interpreted as indicating {different} variability in field goal success probabilities among different players.} In addition, the effect of playing at home, as reflected in the credible intervals of $\beta_{H}^{(C_1)}$, $\beta_{H}^{(C_2)}$, and $\beta_{H}^{(C_3)}$, appears to be around zero for one-point and three-point shots, but slightly positive for made two-point field goals.

\begin{table}[H]
\centering
\footnotesize
\caption{Summary of the approximated posterior distribution of the parameters and hyperparameters of the 1PT, 2PT and 3PT shots scored submodels.}
\label{tab:shots_made_new}
\vspace{0.25cm}
\begin{tabular}{cccccccccc}
\multicolumn{1}{c}{} & \multicolumn{3}{c}{{1PT Made ($k=1$)}} & \multicolumn{3}{c}{{2PT Made ($k=2$)}} & \multicolumn{3}{c}{{3PT Made ($k=3$)}} \\
\cmidrule(rl){2-4} \cmidrule(rl){5-7} \cmidrule(rl){8-10}
 & {Mean} & {Sd} & {0.95CI} & {Mean} & {Sd} & {0.95CI} & {Mean} & {Sd} & {0.95CI}  \\
\midrule
$\beta_{0}^{(C_k)}$   & 0.457 & 0.428 & (-0.333,1.329) & 0.233 & 0.172 & (-0.115,0.559) & -3.434 & 2.892 & (-9.657,1.683) \\
$\beta_{H}^{(C_k)}$   & 0.045 & 0.099 & (-0.148,0.246) & 0.067 & 0.056 & (-0.041,0.175) & -0.060 & 0.132 & (-0.321,0.202)\\
$\beta_{PF}^{(C_k)}$  & 0.452 & 0.518 & (-0.587,1.517) & -0.386 & 0.214 & (-0.798,0.029) & 2.323 & 2.910 & (-2.876,8.598) \\
$\beta_{PG}^{(C_k)}$  & 0.699 & 0.541 & (-0.431,1.674) & -0.274 & 0.229 & (-0.712,0.168) & 2.726 & 2.904 & (-2.421,8.999)\\
$\beta_{SF}^{(C_k)}$  & 1.271 & 0.730 & (-0.300,2.663) & -0.493 & 0.300 & (-1.072,0.123) & 3.145 & 2.923 & (-2.109,9.595)\\
$\beta_{SG}^{(C_k)}$  & 0.924 & 0.585 & (-0.240,2.079) & -0.507 & 0.249 & (-0.998,-0.004) & 2.578 & 2.931 & (-2.578,8.952) \\ [0.6 ex]
${\sigma_{C_k}}$  & 0.475 & 0.239 & (0.175,1.053)  & 0.190 & 0.093 & (0.059,0.416)  & 0.302 & 0.351 & (0.012,1.163) \\
\midrule
\end{tabular}
\end{table}

{Furthermore, bearing in mind that the player's position associated with   the reference category is the \textit{center}, there appears to be a relationship between the probability of
scoring and certain player positions on the pitch. The shots made by the \textit{center} players (in the early 2000s) were mainly two-point shots scored from a close position to the basket. This fact makes that the percentage of success was higher to the ones that are tried from a further positions, which are the most common shots of \textit{power forward} and \textit{shooting forward} players. That explains why the credible intervals of the two point shots of these players compared with the center players are mostly negative ($\beta_{PF}^{(C_2)}$ and $\beta_{SF}^{(C_2)}$). On the other hand, \textit{shooting guards} are generally among the shortest players on the team. They usually attempt shots beyond the three-point line and tend not to play close to the basket, as they can be more easily blocked by opposing players. This is reflected in the credible intervals for \textit{shooting guards} compared with \textit{centers} ($\beta_{SG}^{(C_2)}$ and $\beta_{SG}^{(C_3)}$). Moreover, Table \ref{tab:prob_gt_zero_shots_made}, which presents the posterior probability that each parameter is greater than zero, supports the interpretations discussed above.

\begin{table}[H]
\centering
\footnotesize
\caption{Posterior probability that each parameter is greater than zero for the 1PT, 2PT and 3PT shots made submodels.}
\label{tab:prob_gt_zero_shots_made}
\vspace{0.25cm}
\begin{tabular}{cccc}
\cmidrule(rl){2-4}
 & {1PT Made ($k=1$)} & {2PT Made ($k=2$)} & {3PT Made ($k=3$)} \\
\midrule
$\Pr(\beta_{0}^{(C_k)}>0\, {\,\mid \mathcal D})$   & 0.896 & 0.933 & 0.102 \\
$\Pr(\beta_{H}^{(C_k)}>0\,{\,\mid \mathcal D})$   & 0.667 & 0.884 & 0.319 \\
$\Pr(\beta_{PF}^{(C_k)}>0\,{\,\mid \mathcal D})$  & 0.849 & 0.032 & 0.780 \\
$\Pr(\beta_{PG}^{(C_k)}>0\,{\,\mid \mathcal D})$  & 0.927 & 0.087 & 0.830 \\
$\Pr(\beta_{SF}^{(C_k)}>0\,{\,\mid \mathcal D})$  & 0.961 & 0.043 & 0.873 \\
$\Pr(\beta_{SG}^{(C_k)}>0\,{\,\mid \mathcal D})$  & 0.951 & 0.024 & 0.812 \\
\midrule
\end{tabular}
\end{table}

It should also be noted that we have a sample from the posterior distribution of the random effects associated with each player. By combining all the available information, we can estimate the probability of making any of the three types of shots for any player in a game, according to expression (\ref{eqn:succ_prob}). Specifically, we visualise the success probabilities of two outstanding players: Allen Iverson and Kyle Korver. Figure 4 shows the approximate posterior distributions of the success probabilities for 1PT, 2PT, and 3PT shots for these two players. We observe that, for both players, the free throw is the easiest shot, exhibiting the highest probability of success. This is followed by field goals, with a success probability of around $0.5$ in both cases. Finally, the three-point shot appears, as expected, to have the lowest probability of success. However, at this point, a notable difference emerges between the two players. Kyle Korver clearly shows a much higher probability of successfully making a three-point shot than Allen Iverson.

\begin{figure}[H]
    \centering

    \begin{subfigure}[t]{0.48\textwidth}
        \centering
        \includegraphics[width=\linewidth]{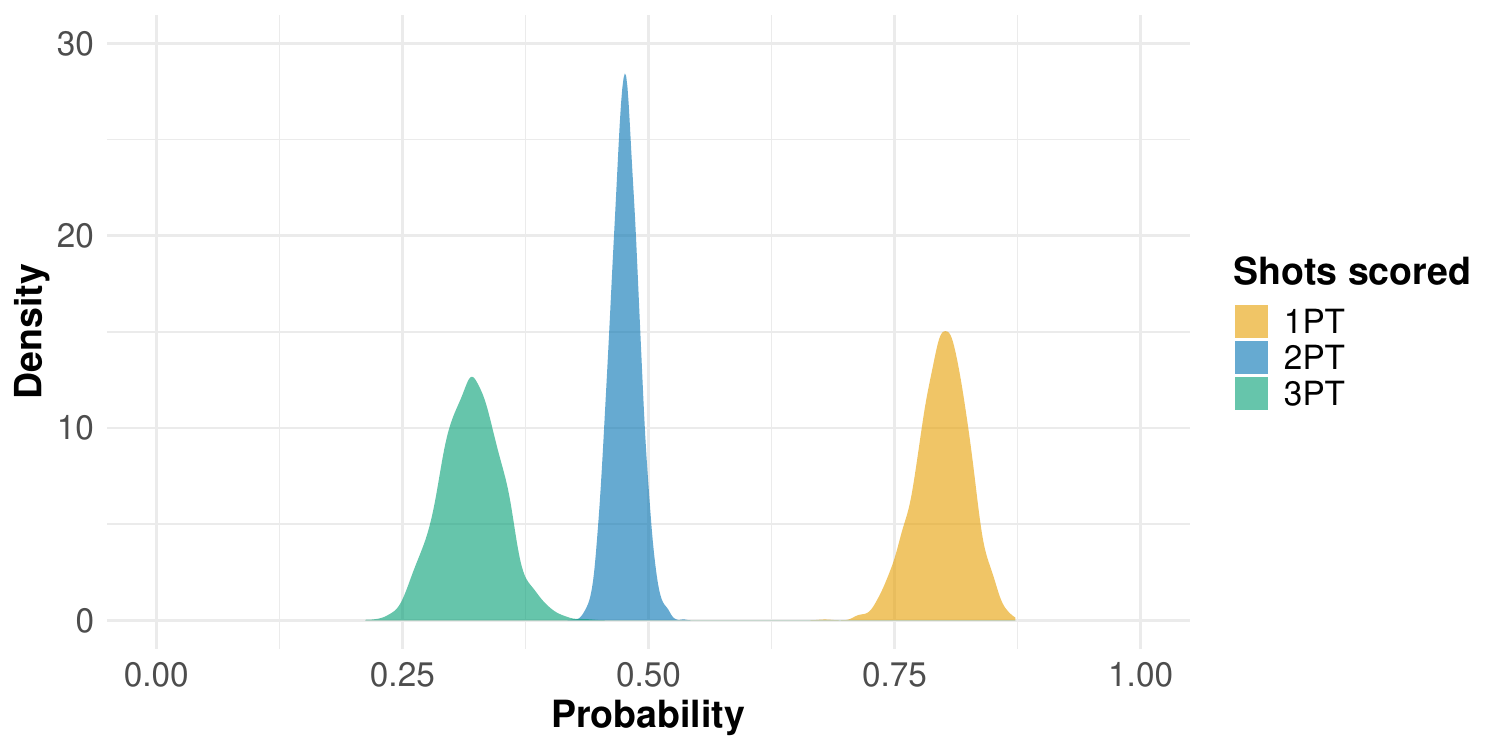}
        \caption{Allen Iverson.}
        \label{fig:prob_success1}
    \end{subfigure}
    \hfill
    \begin{subfigure}[t]{0.48\textwidth}
        \centering
        \includegraphics[width=\linewidth]{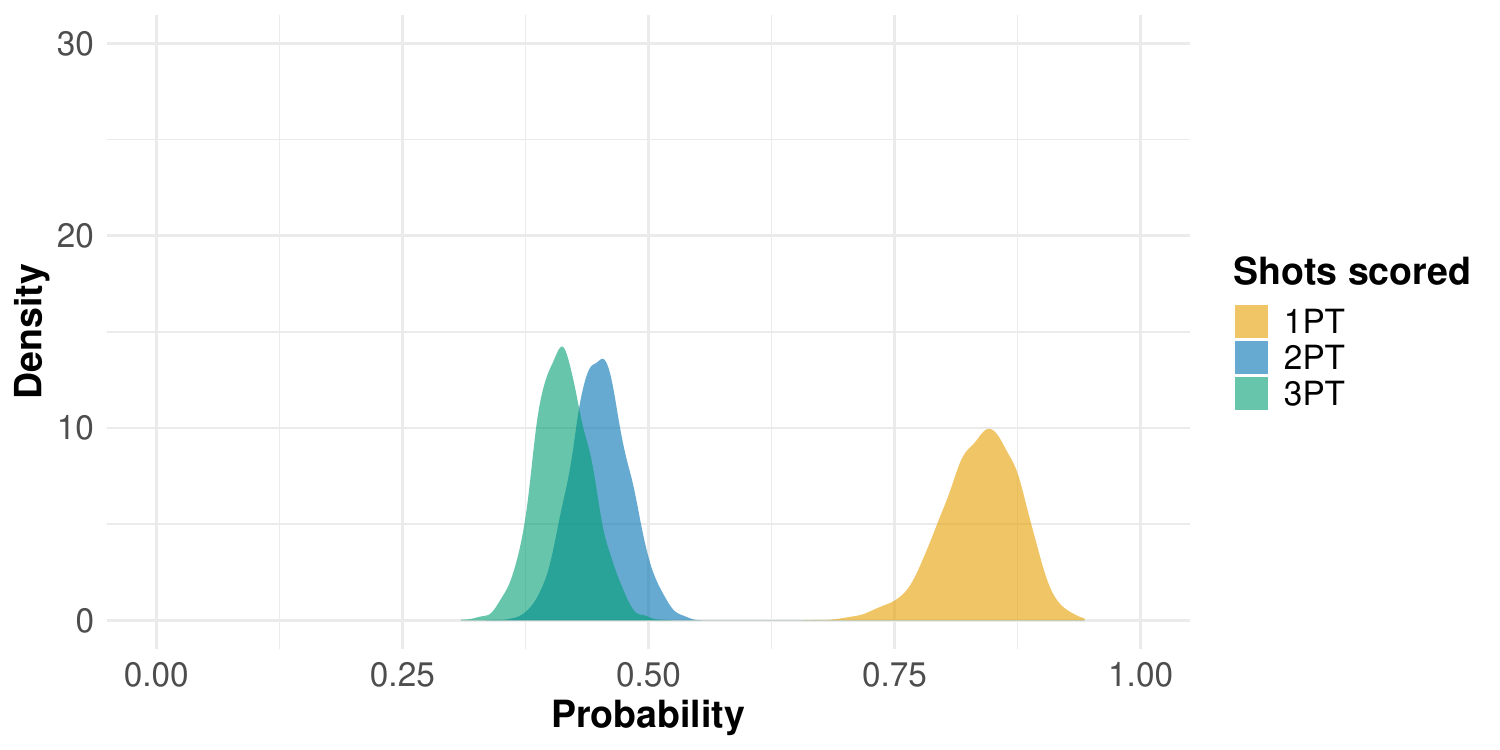}
        \caption{Kyle Korver.}
        \label{fig:prob_success2}
    \end{subfigure}

    \caption{Approximate posterior distributions of the success probabilities
    for 1PT, 2PT, and 3PT shots by Allen Iverson and Kyle Korver.}
    \label{fig:prob_success}
\end{figure}

{Table \ref{tab:shots_attempted} presents a summary of the posterior distribution of the different types of shots attempted. We observe that the credible interval for the three-point attempts of \textit{center} players, $\beta_{0}^{(T_3)}$, is negative. As this position typically operates inside the three-point line, players in this role have fewer opportunities to attempt shots from long range. 
In addition, the parameters $\beta_F^{(T_1)}$, $\beta_F^{(T_2)}$, and $\beta_F^{(T_3)}$ have approximated posterior distributions with positive values. This indicates that the number of fouls received influences the number of free throws, and that the minutes played also affect the volume of field-goal attempts.
In contrast, $\beta_{PG}^{(T_3)}$ is entirely positive, meaning that \textit{point guards} generally attempt more three-point shots than other positions.  In addition, the 95\% credible interval for the \textit{small forward} position in three-point attempts, ($\beta_{SF}^{(T_3)}$), is entirely positive.  This is mainly because the most efficient three-point shooters in those years played mainly as \textit{point guards} or \textit{small forwards}.  }

\begin{table}[H]
\centering
\footnotesize
\caption{Summary of the approximated posterior distribution of the parameters
and hyperparameters of the 1PT, 2PT and 3PT shots  attempted  submodels.}
\label{tab:shots_attempted}
\vspace{0.25cm}
\begin{tabular}{cccccccccc}
\multicolumn{1}{c}{} & \multicolumn{3}{c}{{1PT Attempted  ($k=1$)}} & \multicolumn{3}{c}{{2PT Attempted ($k=2$)}} & \multicolumn{3}{c}{{3PT Attempted ($k=3$)}} \\
\cmidrule(rl){2-4} \cmidrule(rl){5-7} \cmidrule(rl){8-10}
 & {Mean} & {Sd} & {0.95CI} & {Mean} & {Sd} & {0.95CI} & {Mean} & {Sd} & {0.95CI}  \\
\midrule
$\beta_{0}^{(T_k)}$   & 0.246 & 0.830 & (-1.268,2.110) & 0.591 & 0.398 & (-0.196,1.340) & -5.900 & 1.703 & (-9.587,-2.810) \\
$\beta_F^{(T_k)}$       & 0.227 & 0.009 & (0.208,0.245)  & 0.040 & 0.002 & (0.036,0.044)  & 0.030  & 0.005 & (0.021,0.039) \\
$\beta_{PF}^{(T_k)}$  & -0.811 & 1.050 & (-3.061,1.046) & -0.064 & 0.486 & (-1.016,0.919) & 2.471 & 2.076 & (-1.758,6.590) \\
$\beta_{SF}^{(T_k)}$  & -0.305 & 1.383 & (-3.275,2.262) & -0.546 & 0.743 & (-1.997,0.904) & 6.659 & 2.748 & (1.357,12.635) \\
$\beta_{PG}^{(T_k)}$  & -0.333 & 1.085 & (-2.873,1.616) & 0.009 & 0.533 & (-1.061,1.032) & 5.135 & 2.100 & (1.148,9.665) \\
$\beta_{SG}^{(T_k)}$  & -0.330 & 1.153 & (-2.823,1.873) & -0.182 & 0.592 & (-1.369,0.999) & 3.705 & 2.327 & (-0.919,8.593) \\[0.6ex]
$\sigma_{T_k}$   & 1.124 & 0.409 & (0.605,2.112)  & 0.598 & 0.185 & (0.342,1.062)  & 2.194 & 1.003 & (0.947,4.593) \\
\midrule
\end{tabular}
\end{table}

 Table \ref{tab:MP}  refers to the minutes played by each player in a game and reports the posterior distribution  of a common mean of the participation of each  player  in a match, $\mu_{0}^{(M)}$, the autoregressive $\beta_{M}^{(+)}$ regression coefficient associated to the minutes played by each player in the previous game, and the standard deviation $\sigma_{M}$ of the random effect associated to each player.  We observe that the autoregressive coefficient   responsible for turning this {LBN into a dynamic LBN} takes relatively small values. However, the $95\%$ credible interval of the approximated posterior distribution is entirely positive. 

\begin{table}[H]
\centering
\footnotesize
\caption{Summary of the approximate posterior distribution of the parameters and hyperparameters of the minutes played  submodels.}
\label{tab:MP}
\vspace{0.25cm}
\begin{tabular}{cccc}
\multicolumn{1}{c}{} & \multicolumn{3}{c}{} \\
\cmidrule(rl){2-4} 
 & {Mean} & {Sd} & {0.95CI}   \\
\midrule
$\mu_{0}^{(M)}$  & 2.616  & 0.223 & (2.172, 3.004) \\
$\beta_{M}^{(+)}$  & 0.097  & 0.011 & (0.076, 0.119) \\
$\sigma_{M}$        & 0.688   & 0.177 & (0.451, 1.089) \\
\end{tabular}
\end{table}

 Finally,
Table \ref{tab:prob} displays a summary of the posterior distribution of the individual probabilities of playing {a match}. High values can be seen almost all the players which are even higher for the stars of the team (Iverson, Iguodala and Webber). Less dominance can be seen in the \textit{centers} (Dalmebert, Randolph and Hunter) which had the same probability of playing. It is also interesting to see that Korver and Salmons have a high probability, that is because they were seen as the young promises of the team.

\begin{table}[H]
\centering
\footnotesize
\caption{Summary of the approximated posterior distribution of the individual probabilities of playing.}
\label{tab:prob}
\vspace{0.25cm}
\begin{tabular}{lccc}
 \multicolumn{3}{c}{} & \\
\cmidrule(rl){2-4} 
{Player} & {Mean} & {Sd} & {0.95CI}  \\
\midrule
Allen Iverson & 0.870 & 0.036 & (0.792, 0.932) \\
Andre Iguodala & 0.988 & 0.012 & (0.958, 1.000)\\
Chris Webber & 0.905 & 0.031 & (0.835, 0.957) \\
John Salmons & 0.988 & 0.012 & (0.958, 1.000)   \\
Kevin Ollie   & 0.845 & 0.039 & (0.764, 0.914)   \\
Kyle Korver  & 0.988 & 0.012 & (0.959, 1.000)   \\
Lee Nailon   & 0.273 & 0.047 & (0.188, 0.372)  \\
Lou Williams  & 0.369 & 0.054 & (0.272, 0.483)   \\
Matt Barnes  & 0.606 & 0.053 & (0.501, 0.706)  \\
Michael Bradley& 0.561 & 0.053 & (0.454, 0.664)  \\
Samuel Dalembert   & 0.797 & 0.044 & (0.709, 0.878) \\
Shavlik Randolph & 0.690 & 0.051 & (0.585, 0.788) \\
Steven Hunter & 0.833 & 0.040 & (0.750, 0.906)  \\
\end{tabular}
\end{table}

\noindent
 
\subsection{Prediction}
A powerful feature of BNs is their capacity to compute predictions by systematically applying {the total law of probability} within their structured probabilistic framework.  This architecture allows not only the prediction of variables conditioned
by others that influence them probabilistically, but also  flexible predictions of
any variables given new evidence about anothers.
This versatility, in turn, facilitates multidirectional reasoning, allowing {prediction} from causes to effects and from effects back to causes. In the Bayesian framework, we use the joint posterior predictive distribution of {the possible participation, $\boldsymbol y_*^{(A)}$, the number $\boldsymbol y_*^{(M)}$ of minutes played, the number $\boldsymbol y_*^{(F)}$ of fouls drawn, and the number of shots thrown and scored, $\boldsymbol y_*^{(T)}$ and  $\boldsymbol y_*^{(C)}$, respectively,  of each player of the team in a  new hypothetical match}   defined  as
\begin{align}
\label{predictive}
f(\boldsymbol{y}_*^{(C)}, & \,\boldsymbol{y}_*^{(T)}, \boldsymbol{y}_*^{(F)},   \boldsymbol{y}_*^{(M)}, \boldsymbol{y}_*^{(A)} \mid \mathcal{D}) =   \nonumber \\
& \mbox{$\int$}\, f(\boldsymbol{y}_*^{(C)}, \boldsymbol{y}_*^{(T)}, \boldsymbol{y}_*^{(F)}, \boldsymbol{y}_*^{(M)}, \boldsymbol{y}_*^{(A)} \mid \boldsymbol{\theta},\boldsymbol{\phi}) \,\pi(\boldsymbol{\theta},\boldsymbol{\phi}\mid \mathcal{D}) \, \mbox{d} (\boldsymbol{\theta}, \boldsymbol \phi).
\end{align}
From the complete posterior predictive distribution, it is possible to compute different types of {posterior predictive marginals and  conditionals}   outputs. The analysis may focus on those predictions that are of particular interest for the model. For instance, one may predict how many points in total, counting those scored from free throws, two-point shots and three-point shots, will the player score in the next match in the case the player would play more {or less} than    a {given number of minutes}. Figure \ref{fig:pred1} presents an approximation of this predictive distribution   for Allen Iverson and Kyle Korver {in the event that they played for more than 30 minutes}. The results show clearly that the predicted points for Allen Iverson are higher than those for Kyle Korver.

\begin{figure}[H]
	\centering
 \includegraphics[width=110mm]{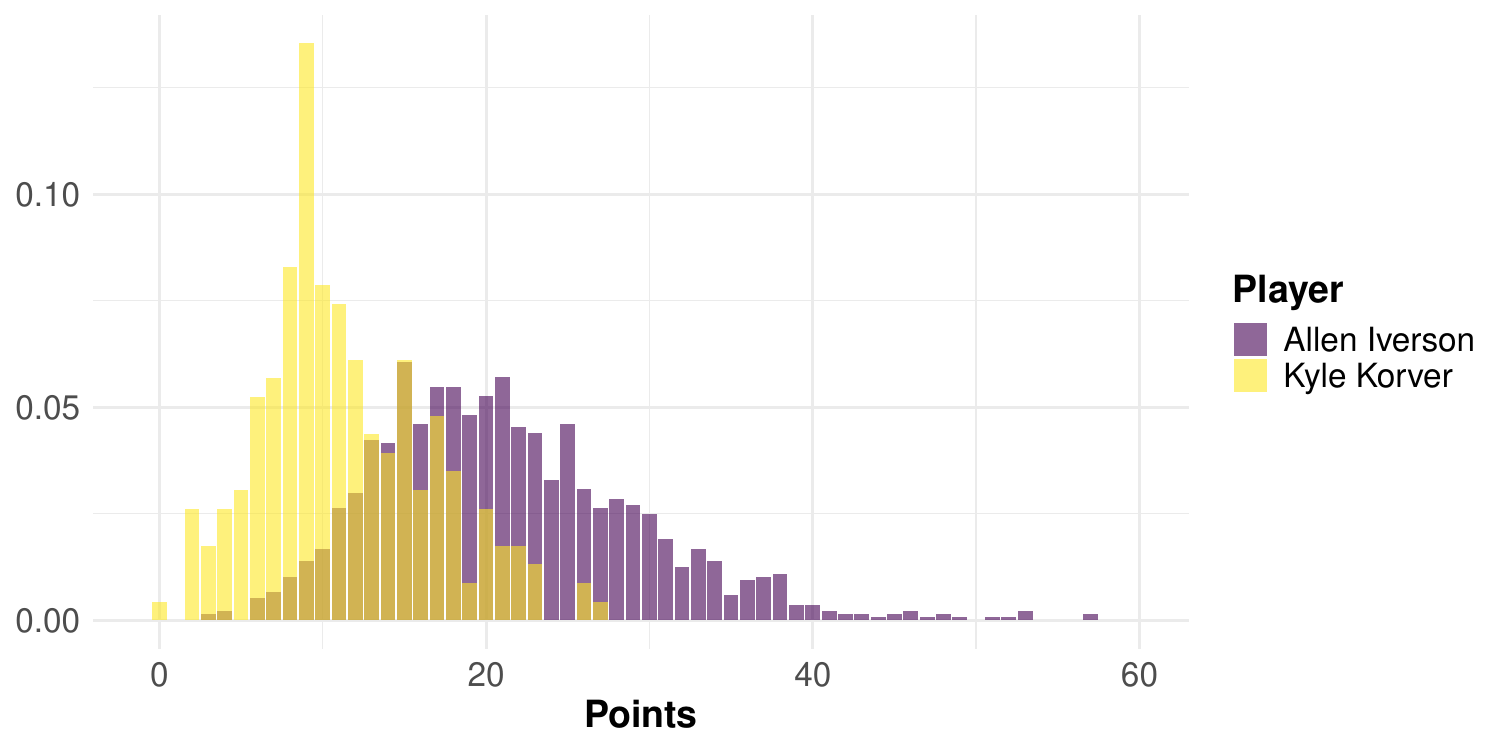}
	 	\caption{Approximate predictive distribution of the total number of points scored (1PT + 2PT+ 3PT) by the players Allen Iverson and Kyle Korver  in a new match, assuming they play more than 30 minutes.} \label{fig:pred1}
\end{figure}

{We can also approach the problem the other way round. That is, knowing that a particular player has scored  {more or less than a certain number of points}, we could predict how many minutes he has played during the match. Figure \ref{fig:pred2} shows the approximate posterior predictive distribution of the minutes played given that the player, Allen Iverson or Kyle Korver, has scored 10 points or fewer in a match. It can be observed that the posterior predictive distribution of the minutes played, given the points scored, is a mixture probability distribution, as a consequence of how the model is defined in expression (\ref{eqn:ZIP}). In this sense, it can be observed in Figure \ref{fig:pred2} that if Allen Iverson scores fewer than 10 points in a game, this is likely because he did not play or played very few minutes. This does not occur with Kyle Korver, as can be seen in Figure \ref{fig:pred2}.}

\begin{figure}[H]
	\centering
 \includegraphics[width=110mm]{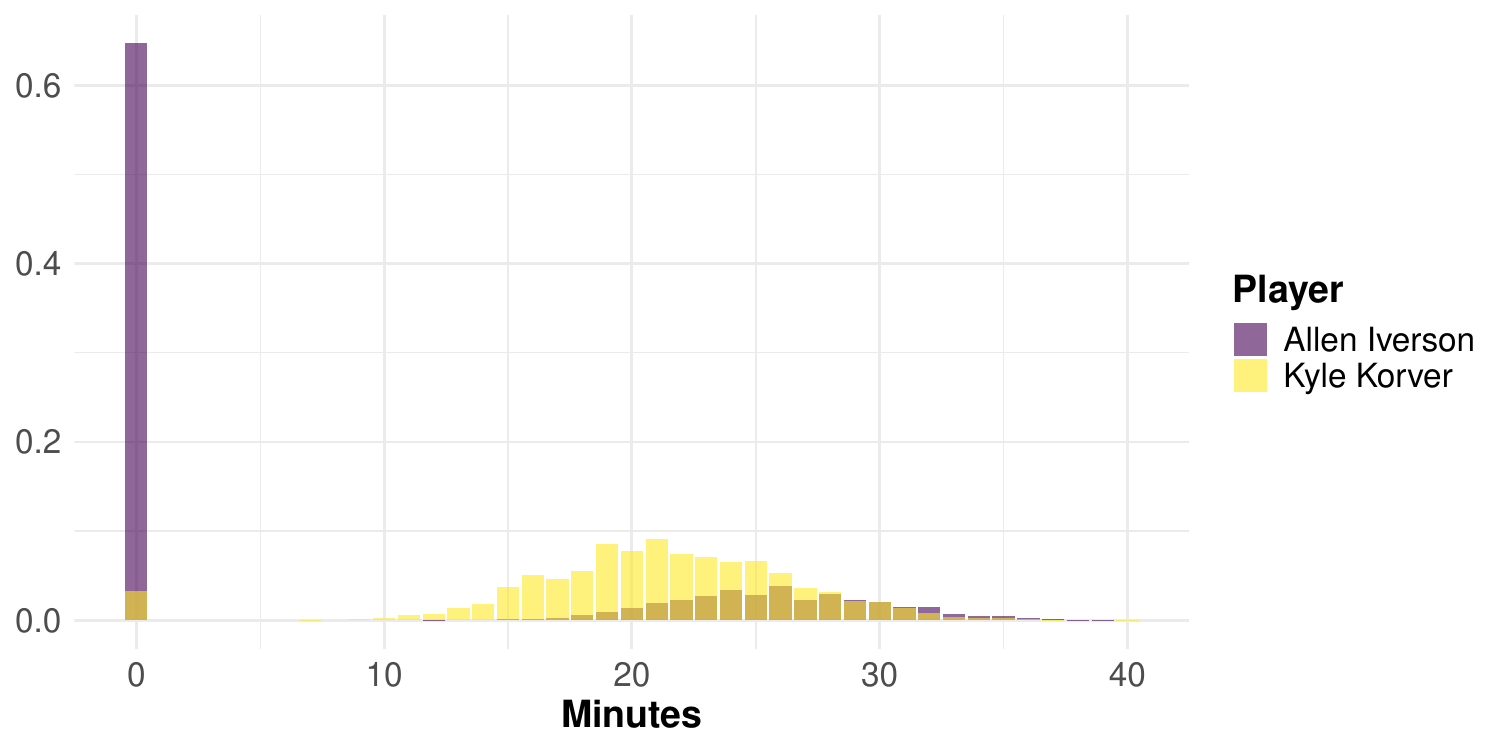}
	 	\caption{Approximate predictive distribution of the total number of minutes played by Allen Iverson and Kyle Korver in a new match assuming they score a total of  10 points or fewer.} \label{fig:pred2}
\end{figure}

\section{Conclusions, limitations and future research lines}

The main contribution of this work is to develop a longitudinal analysis of basketball team performance within a Bayesian graphical modelling framework. {In particular, we discuss three basic, generic longitudinal models for evaluating the performance and effectiveness of players in a basketball team over the course of an entire league season: a static longitudinal Bayesian network, a dynamic longitudinal Bayesian network incorporating an autoregressive dependence structure between  {succesive matches}, and a dynamic longitudinal Bayesian network based on a hidden Markov structure. The proposed models can easily be adapted to other leagues beyond the NBA and to updated schedules, while allowing for the straightforward integration of additional relevant nodes and covariates. }

In the specific case of the Philadelphia 76ers, the results showed that the autoregressive model, referred to as the dynamic LBN model, provided the best predictive performance, as indicated by its lowest WAIC value. Among other findings, the proposed framework enabled us to predict both the points scored and the minutes played by all the players of the team. At the player level, the posterior predictive distributions for Allen Iverson and Kyle Korver indicate that, when Iverson plays more than 30 minutes, he is expected to score considerably more points than Korver. They also suggest that  Iverson points total below 10 is associated with a high posterior predictive probability of him not having played in that game. This pattern is not observed for Kyle Korver.

Given that automatic graph structure learning constitutes a highly complex combinatorial problem, particularly in the presence of latent nodes, its investigation has been deliberately placed beyond the scope of this study. Accordingly, the network structure is assumed to be specified by the researcher or treated as known.

This modelling perspective naturally suggests several directions for future research. A first extension would be to allow for multivariate nodes, enabling a
single node to represent a set of correlated variables that evolve jointly
over time, such as the numbers of free throws, two-point field goals, and
three-point field goals made in each match. An especially promising line of work concerns the incorporation of spatial or spatio-temporal nodes, which would make it possible to capture spatial dependencies and their interaction with longitudinal dynamics. In addition, further research should address approximate approaches to structural learning, as well as extensions of the framework to unbalanced data and irregular measurement times. Altogether, these developments would complement the framework proposed here and considerably broaden its applicability to a wider range of complex longitudinal phenomena.

\section*{Acknowledgements}
This work is part of the project PID2022-136455NB-I00, funded by Ministerio de Ciencia, Innovación y Universidades of Spain (MCIN/AEI/10.13039/501100011033/ FEDER, UE) and the European Regional Development Fund.

\bibliographystyle{apalike}
\bibliography{biblio}
\end{document}